\documentclass[twocolumn]{aa}

\usepackage{graphicx}
\usepackage{microtype}
\usepackage{placeins}
\usepackage{txfonts}
\usepackage{xspace}
\usepackage{multicol}
\usepackage{multirow}
\usepackage{color}
\usepackage{url}
\usepackage{hyperref}
\usepackage[flushleft]{threeparttable}
\newcommand {\bc}{\begin {center}}
\newcommand {\ec}{\end {center}}
\newcommand {\be}{\begin {equation}}
\newcommand {\ee}{\end {equation}}
\newcommand {\beq}{\begin {eqnarray}}
\newcommand {\eeq}{\end {eqnarray}}
\newcommand {\ergs}{{\rm erg\ \rm s^{-1}}}

\usepackage{graphicx}
\usepackage{txfonts}

\begin{document}

\title{Continuous evolution of the polarization properties in the transient X-ray pulsar RX J0440.9+4431/LS V +44 17}
\titlerunning{RX J0440.9+4431 Polarization Evolution}
\author
{
Q. C. Zhao\inst{1,2}
\and L. Tao\inst{1}\thanks{E-mail: taolian@ihep.ac.cn}
\and Sergey~S.~Tsygankov\inst{3}\thanks{E-mail: sergey.tsygankov@utu.fi}
\and Alexander~A.~Mushtukov\inst{4}
\and H. Feng\inst{1}
\and M. Y. Ge\inst{1}
\and H. C. Li\inst{5}
\and S. N. Zhang\inst{1}
\and L. Zhang\inst{1}
}
\institute{Key Laboratory of Particle Astrophysics, Institute of High Energy Physics, Chinese Academy of Sciences, Beijing 100049, China
\and University of Chinese Academy of Sciences, Chinese Academy of Sciences, Beijing 100049, China
\and Department of Physics and Astronomy, FI-20014 University of Turku,  Finland
\and Astrophysics, Department of Physics, University of Oxford, Denys Wilkinson Building, Keble Road, Oxford OX1 3RH, UK
\and Department of Astronomy, University of Geneva, 16 Chemin d’Ecogia, Versoix, CH-1290, Switzerland
 }
\abstract{We present a detailed time-resolved and phase-resolved polarimetric analysis of the transient X-ray pulsar RX J0440.9+4431/LS V +44 17, using data from the Imaging X-ray Polarimetry Explorer (IXPE) during the 2023 giant outburst. We conducted a time-resolved analysis by dividing the data into several intervals for each observation. This analysis reveals a continuous rotation of the phase-averaged polarization angle (PA) across the observations performed during the super-critical and sub-critical regimes. To investigate the origin of the PA rotation, we performed a pulse phase-resolved polarimetric analysis over four time intervals, each spanning approximately three days. Applying the rotating vector model (RVM), the geometric parameters of the system were determined for each interval. Despite the short time gap of just $\sim$ 20 days, we observed significant variation in the RVM parameters between the first interval and the subsequent three, indicating the presence of an additional polarized component alongside the RVM component. Using a two-polarized component model with the assumption that this additional component remains constant across pulse phases, we calculated the phase-averaged PA and polarized flux of both the variable and constant components. The phase-averaged PA of each component remained relatively stable over time, but the polarized flux of the constant component decreased, while that of the variable component increased. The observed rotation of the PA is attributed to the gradual shift in the polarized flux ratio between the two components and is not directly related to the different accretion regimes.}

\keywords{magnetic fields – methods: observational – polarization – pulsars: individual: RX J0440.9+4431 – stars: neutron – X-rays: binaries}

\maketitle
%
%

\section{Introduction}
Accreting X-ray pulsars (XRPs) are neutron stars (NS) in binary systems that accrete material from a companion star via either an accretion disk or a stellar wind. The surface magnetic field strength of the NS in XRPs can reach values as high as \(\sim 10^{12}\) to \(10^{13}\) G. This strong magnetic field channels the infalling plasma onto the magnetic poles, leading to the production of pulsating X-ray emission. The observational properties of XRPs are influenced by a multitude of factors, such as the geometry of the emission region (e.g., hot spot, mound, or accretion column), the observer’s viewing angle in relation to the magnetic pole and rotational axes of the pulsars, and, fundamentally, the accretion rate and intricate emission mechanisms within the strong magnetic field. Exploring these systems is pivotal for understanding the interaction between X-ray radiation and strongly magnetized plasma \citep[see a recent review by][]{Mushtukov_Tsygankov_review}.

The apparent luminosity of XRPs spans several orders of magnitude, ranging from approximately \(10^{32}\) erg s\(^{-1}\) to \(10^{41}\) erg s\(^{-1}\). The most luminous XRPs belong to the class of pulsating ultraluminous X-ray sources (ULXs; see, e.g., \citealt{2021AstBu..76....6F,2023NewAR..9601672K}).
The geometry of the emitting region depends on the luminosity or more fundamentally, the mass accretion rate. 
At high accretion rates (in the super-critical regime), a radiation-dominated shock forms and an accretion column is built up above the surface \citep{Basko_etal_1976,2015MNRAS.447.1847M}. In this scenario, the majority of the radiation diffuses within the accretion column and escapes through its walls, producing a characteristic "fan" beam pattern. At low mass accretion rates (in the sub-critical regime), material reaches the NS’s surface via a gas-mediated shock, resulting in the formation of a characteristic "pencil" beam pattern. At intermediate luminosity levels (around \(10^{37}\) erg s\(^{-1}\)), the beam pattern can become complex, potentially exhibiting a hybrid of both "pencil" and "fan" beam configurations \citep{Becker_etal_2012,2015MNRAS.448.2175L}. Therefore, variations in the beam pattern can lead to alterations in pulse profiles, which can be used to identify the critical luminosity \citep[see, e.g.,][]{Doroshenko_etal_2020, Rai_etal_2021, Wang_etal_2022, Hu_etal_2023, Salganik_etal_2023, Li_etal_2024}. The non-isotropic emission of XRPs, combined with their rotation, can result in a discrepancy between the intrinsic and observed X-ray luminosity. However, this difference is expected to be relatively minor (\(<20\%\)) at luminosities \(\lesssim 10^{39} \, \mathrm{erg \, s^{-1}}\) \citep{2024MNRAS.527.5374M}. This discrepancy becomes more pronounced at the higher luminosities typical of the brightest XRPs, where intense accretion can trigger strong outflows from the accretion disc \citep{2009MNRAS.393L..41K,2023MNRAS.518.5457M}.

To measure the intrinsic beam function of an XRP, one needs first to determine the orientation of the NS relative to the observer. X-ray polarization has already been demonstrated to be a powerful tool for probing the geometric configuration of XRPs. Several sources have been observed with the Imaging X-ray Polarimetry Explorer (IXPE), providing valuable insights into their polarization properties. These sources include Her X-1 \citep{Doroshenko_etal_2022_herx-1,Garg_herx-1,Zhao_herx-1,Heyl_herX-1}, Cen X-3 \citep{Tsygankov_etal_2022_cenx-3}, GRO J1008--57 \citep{Tsygankov_groj1008}, 4U 1626--67 \citep{Marshall_4U1627}, X Persei \citep{Mushtukov_xpeisei}, Vela X-1 \citep{Forsblom_velaX-1}, EXO 2030+375 \citep{Malacaria_exo2030}, GX 301--2 \citep{Suleimanov_gx301}, RX J0440.9+44331 / LS V +44 17 \citep{Victor_etal_2023}, Swift J0243.6+6124 \citep{SwiftJ0243_Majumder,SwiftJ0243_Poutanen}, and SMC X-1 \citep{SMCX-1_Forsblom}. Most of these sources were observed in the sub-critical state, revealing polarization degrees (PD) significantly lower than theoretical predictions \citep{Meszaros_1988,Caiazzo_polarization_model}. Additionally, 1A 0535+262 exhibited a non-detection of polarization in the super-critical state, with a 99\% confidence level upper limit of 34\%, as measured by PolarLight in the 3–8 keV range \citep{Polarlight, Long_0535}. These findings suggest that low PD is a general characteristic of XRPs, regardless of luminosity levels.

The most intriguing sources in this class for polarimetric studies are transient XRPs, where the evolution of the geometric and physical parameters of the emission region can be examined as a function of the mass accretion rate. One of such XRPs, RX J0440.9+4431 (henceforth, RX J0440), was observed by IXPE in different states during a giant outburst from the source in 2022--2023 \citep{Nakajima_LSV44,Pay_LSV44}. During this outburst, transitions between the sub-critical and super-critical regimes were identified by \citet{Salganik_etal_2023} on MJD \(\sim\)59971 and \(\sim\)MJD 59995, based on the evolution of the pulse profiles. This transition was further confirmed by \citet{Li_etal_2024} using high-energy pulse profiles from the Insight-HXMT data. Spectral parameters obtained from NICER also support the presence of this transition \citep{Mandal_etal_lsv44}. The cyclotron line is not detected in the broadband energy spectra, which may be attributed to the high magnetic field strength, estimated to be approximately \(10^{13} \, \mathrm{G}\) \citep{Salganik_etal_2023,Li_qpq_2024}. \citet{Victor_etal_2023} conducted a phase-resolved polarimetric analysis of two IXPE observations, carried out in super-critical and sub-critical states, revealing a significant difference in the phase-resolved modulations of PD and PA between the two states. They proposed that the observed difference was due to the presence of a constant (unpulsed) polarized component. After subtracting this constant component, the phase-resolved modulations of PA in two observations could be fitted using the same set of the pulsar’s geometric parameters with a rotating vector model (RVM). \citet{Victor_etal_2023} suggested that the constant polarized component may originate from scattering in the disk wind. A similar mechanism is also inferred to be responsible for PA rotation in the weakly magnetized NS GX 13+1 \citet{Bobrikova_GX_13+1}.

In this paper, we extend the investigation of RX J0440's polarization properties by conducting a time-resolved analysis of both IXPE observations. By segmenting the data at a higher time resolution, we are able to track the evolution of the polarization properties with greater precision. Furthermore, applying a two-polarized component model \citep{Victor_etal_2023} allows us to study the evolution of the polarization properties of the two components as a function of both time and luminosity. This paper is organized as follows: in Sect.~\ref{sec:sec2}, we describe the observations and data reduction methods. The results are presented in Sect.~\ref{sec:sec3} and discussed in Sect.~\ref{sec:sec4}.

\begin{figure}
    \centering
    \includegraphics[width=0.5\textwidth]{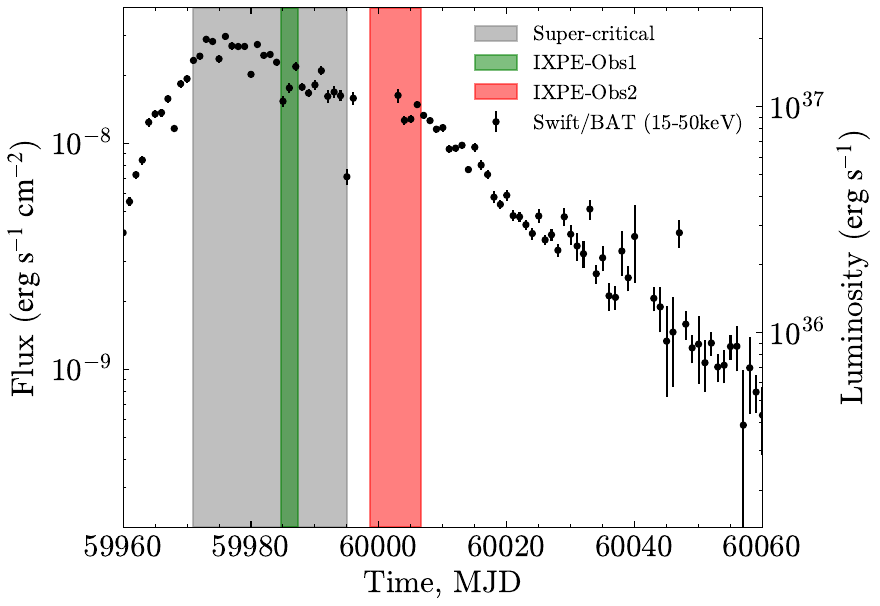}
    \caption{Light curve of RX J0440 during the 2022–2023 outburst. The black points represent Swift/BAT (15–50 keV) data, while the gray strip marks the super-critical regime, as identified by \citet{Salganik_etal_2023} and \citet{Li_etal_2024}. The green and red strips indicate the first and second IXPE observations, respectively. Luminosity values were calculated assuming a distance of 2.4 kpc to the source \citep{RXJ0440_Distance}.
    }
    \label{fig:lightcurve}
\end{figure}
\section{Observations and data reduction} \label{sec:sec2}

IXPE, launched on December 9, 2021, is equipped with three grazing incidence telescopes and gas pixel detectors, enabling polarimetric measurements in the 2–8 keV energy range \citep{Soffitta_etal_2021,Weisskopf_ixpe}. IXPE conducted two observations of RX J0440, as shown in Fig.~\ref{fig:lightcurve}. The first observation occurred during the super-critical regime, while the second was carried out during the sub-critical regime.

We started our analysis using the level-2 data products, which were further processed with {\sc ixpeobssim v31.0.1} \citep{Baldini_etal_2022}.\footnote{\url{https://ixpeobssim.readthedocs.io/en/latest/}} For the analysis, we selected a circular  region around the source with a radius of 100 arcseconds. Background subtraction was not applied, as its contribution was deemed negligible \citep{Di_Marco_etal_2023}. Arrival time corrections were performed using the \texttt{barycorr} tool from the \texttt{HEASoft package v6.32.1}. No binary correction was applied, as the system's orbital parameters were unavailable at the time of writing. However, we note that the orbital period of RX J0440 is approximately 150 days \citep{Tsygankov_RXJ0440_orbit_period,Ferrigno_RXJ0440_orbit_period}, which is much longer than the duration of the analyzed observations. Thus, the results are not significantly affected by orbital motion. We generated polarization cubes with the \texttt{PCUBE} algorithm and the Stokes \textit{I}, \textit{Q}, and \textit{U} spectra using the \texttt{PHA1}, \texttt{PHA1Q}, and \texttt{PHA1U} algorithms, respectively. The \textit{I} spectra were grouped to ensure a minimum of 30 counts per bin, while a constant energy binning of 0.2 keV was applied to the Stokes \textit{Q} and \textit{U} spectra. Throughout this paper, uncertainties are reported at the 68\% confidence level unless otherwise stated.

\begin{figure}
    \centering
    \includegraphics[width=0.45\textwidth]{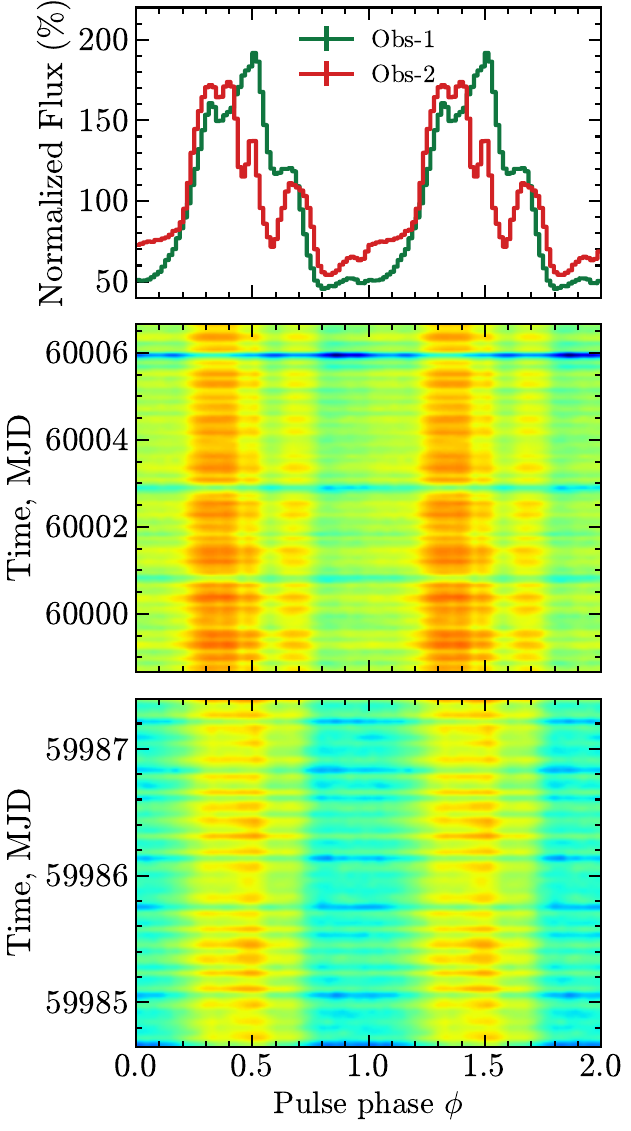}
    \caption{Phase-aligned pulse profiles and phaseograms of RX J0440, based on IXPE data in the 2--8 keV energy band, for Obs-1 and Obs-2, shown in green and red, respectively.\\
    }
    \label{fig:profiles}
\end{figure}

\begin{figure*}[tb]
    \centering
    \includegraphics[width=0.95\textwidth] {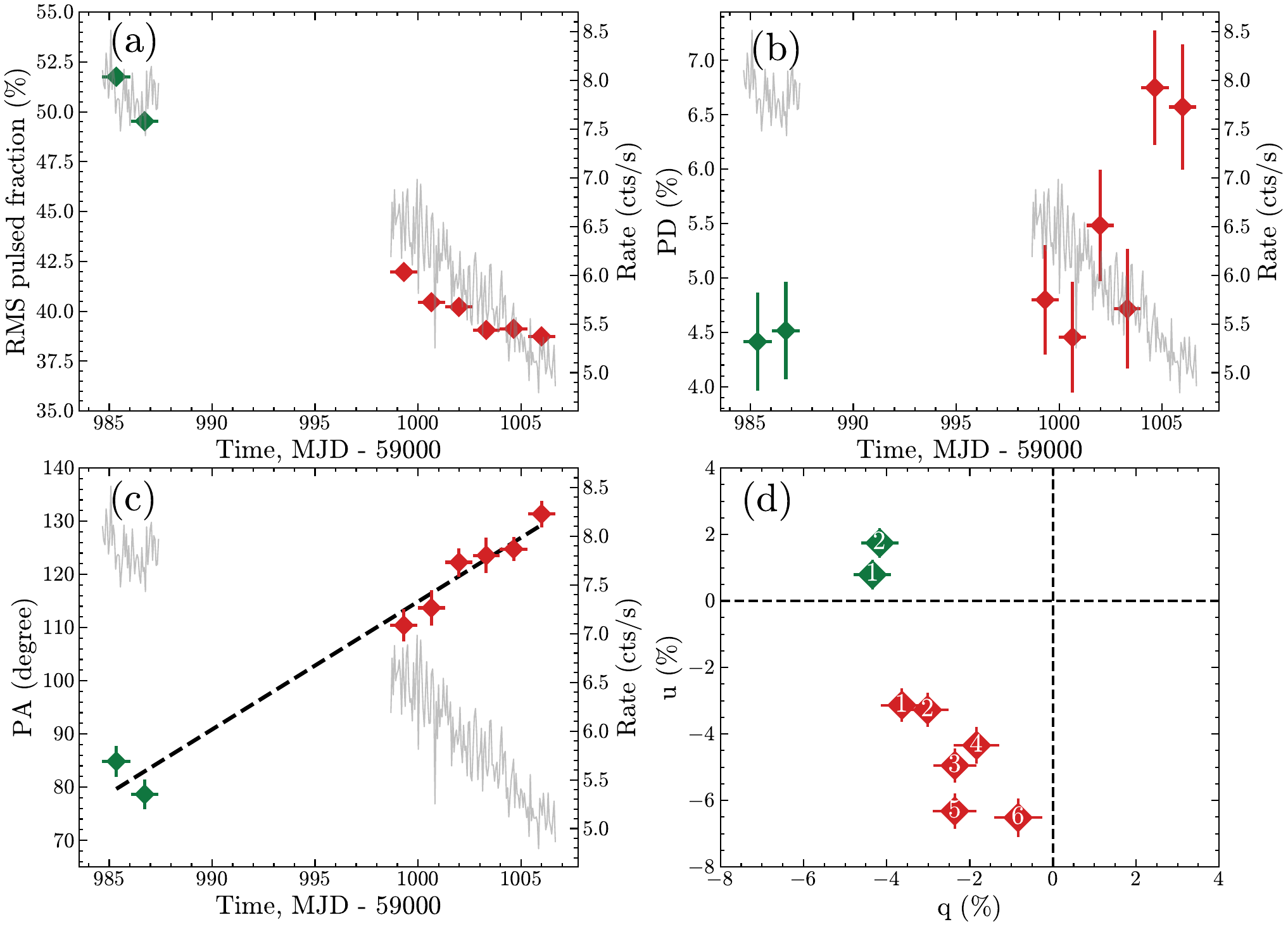}
    
    \caption{Panel (a): time dependence of the RMS pulsed fraction. Panels (b): variations of the phase-averaged PD (2--8 keV) with time. Panel (c): same as panel (b), but for PA. The green and red points represent the first and second observations, respectively. The black dashed line in panel (c) indicates the linear fit for the PA evolution with the slope of \(2.40 \pm 0.16\) deg day$^{-1}$. The gray color represents the DU1 light curve of RX J0440 based on IXPE data in the 2--8 keV band with 5000\,s bin size. Panel (d): the normalized Stokes parameters $q=Q/I$ and $u=U/I$. 
    }
    \label{fig:time_resolved}
\end{figure*}

\section{Results and analysis} \label{sec:sec3}
\subsection{Timing analysis}
\label{sec:sec3.1}

The pulse phases were calculated using the ephemerides listed in Table 1 of \citet{Victor_etal_2023}. Fig.~\ref{fig:profiles} presents the pulse profiles in the 2--8 keV energy band for each IXPE observation, along with the corresponding phaseograms. The flux values have been normalized to the average flux. Notably, the pulse profiles between the two observations exhibit significant differences.

We conducted a time-resolved analysis to examine the evolution of the polarization (see Sect. \ref{sec:sec3.2}) and timing properties over time. The observations were divided into several intervals, with two intervals for Obs. 1 and six for Obs. 2, ensuring that each interval had approximately equivalent statistics. We also calculated the RMS pulsed fraction for each interval using the following formula \citep{Wilson-Hodge_RMS_pulsed_fraction}:
\begin{equation}
	{f}_{\rm rms} = \frac{\left(\sum\nolimits_{i=1}^N(r_i-\bar{r})^2 /N\right)^{1/2}}{\bar{r}}
\end{equation}
The RMS pulsed fraction shows a decreasing trend over time (see Fig.~\ref{fig:time_resolved}a).

\begin{table}[h!]
    \centering
    \resizebox{0.45\textwidth}{!}{%
    \begin{tabular}{cccc} \hline
        Observation  & MJD & PD (\%) & PA (degree) \\ \hline
        02250401   & 59984.647-59986.040 & $4.41\pm0.45$ & $84.8\pm2.9$ \\ 
          & 59986.047-59987.396 & $4.52\pm0.45$ & $78.6\pm2.8$ \\ \hline
     02250501   & 59998.657-59999.974 & $4.80\pm0.50$ & $110.4\pm3.0$ \\
          & 59999.993-60001.311 & $4.46\pm0.51$ & $113.7\pm3.3$\\
          & 60001.331-60002.648 & $5.48\pm0.51$ & $122.3\pm2.7$\\
          & 60002.668-60003.966 & $4.72\pm0.55$ & $123.5\pm3.3$\\
          & 60003.973-60005.323 & $6.48\pm0.53$ & $124.8\pm2.2$\\ 
          & 60005.345-60006.661 & $6.57\pm0.58$ & $131.3\pm2.5$\\ 
         \hline
    \end{tabular}%
    }
   \caption{Pulse phase-averaged PD and PA in different time intervals.}
   \label{tab:phase-averaged evolution}
\end{table}

\subsection{Time-resolved polarimetric analysis} \label{sec:sec3.2}

First, we performed a model-independent phase-averaged polarimetric analysis for Obs. 1 and Obs. 2, using the \texttt{PCUBE} algorithm in \texttt{XPBIN}. The results yielded a PD of \(4.3\% \pm 0.3\%\) at a PA of \(82^{\circ} \pm 2^{\circ}\) for Obs. 1, and a PD of \(5.3\% \pm 0.2\%\) at a PA of \(121^{\circ} \pm 1^{\circ}\) for Obs. 2, consistent with \citet{Victor_etal_2023}. We then performed a spectro-polarimetric analysis by fitting the Stokes \textit{I}, \textit{Q}, and \textit{U} spectra in \texttt{XSPEC} \citep{Arnaud_xspec}, using the \texttt{Constant*Polconst*Tbabs*Comptt} model. This yielded a PD of \(4.3\% \pm 0.2\%\) at a PA of \(79^{\circ} \pm 2^{\circ}\) for Obs. 1, and a PD of \(4.9\% \pm 0.2\%\) at a PA of \(121^{\circ} \pm 1^{\circ}\) for Obs. 2, consistent with the model-independent results.

To investigate the time evolution of the polarization properties, we divided the observations into multiple intervals (see Sect.~\ref{sec:sec3.1}). This analysis reveals a continuous evolution of the PA over time, as shown in Fig.~\ref{fig:time_resolved} and detailed in Table~\ref{tab:phase-averaged evolution}. A linear fit of PA versus time yields a slope of \(2.40 \pm 0.16\) deg day$^{-1}$. Plotting the normalized Stokes parameters in the \(q\)-\(u\) plane, as shown in panel (d) of Fig.~\ref{fig:time_resolved}, demonstrates a clear rotation of the PA with time, accompanied by an increase in the PD from approximately 4\% to 7\%.

\subsection{Phase-resolved polarimetric analysis}

\begin{figure*}[tb]
    \centering
    \includegraphics[width=0.95\textwidth] {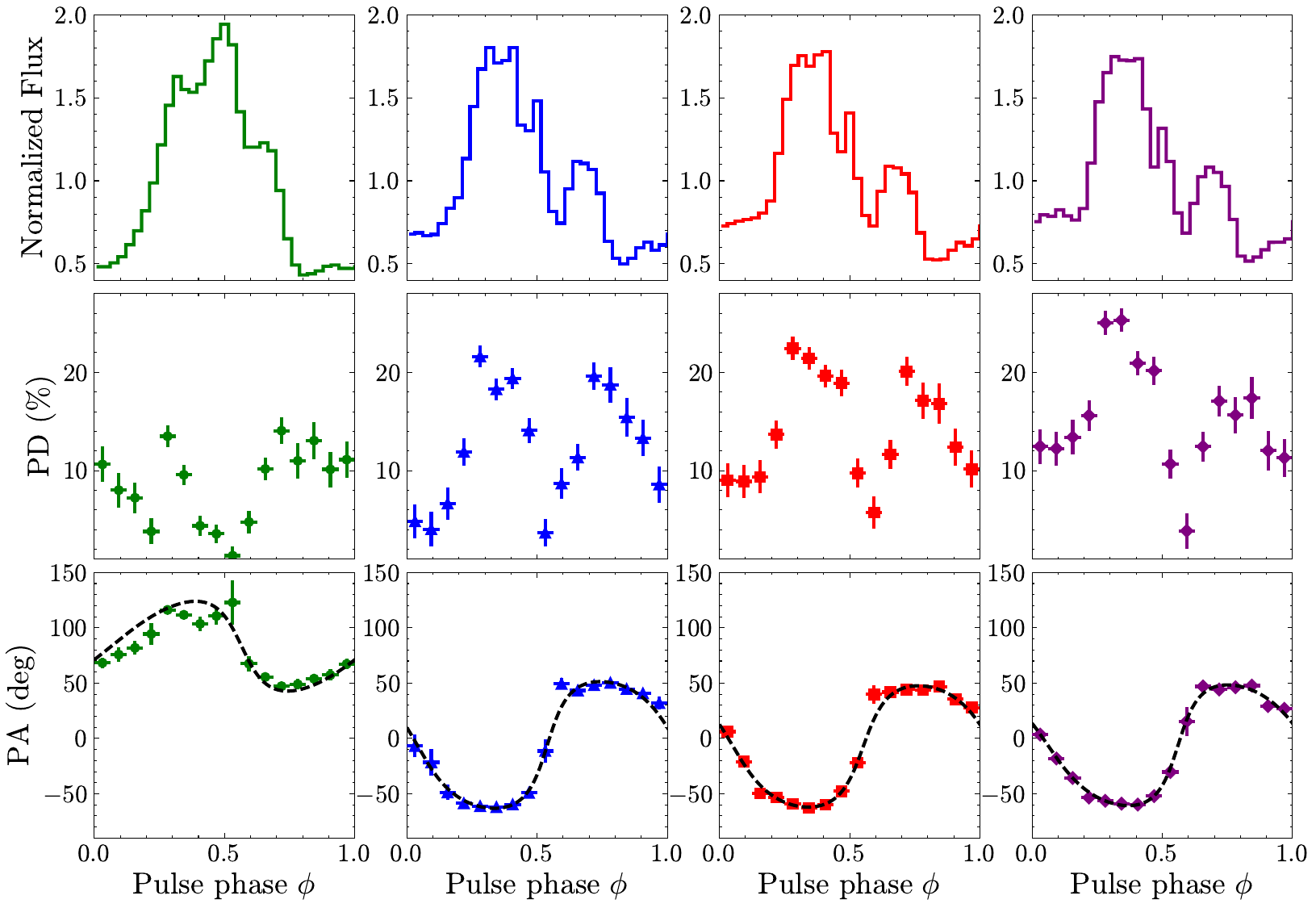}

    \caption{Pulse phase-resolved results of time-resolved polarimetric analysis (2--8 keV). Top panels: pulse profiles. Middle panels: the PD variations with the pulse phase. Bottom panels: same as middle panels, but for the PA. The interval-1 (Obs. 1), interval-2 (Obs. 2), interval-3 (Obs. 2) and interval-4 (Obs. 2) are color encoded with green, blue, red and purple colors. The black dashed lines represent the single-component RVM fit. 
    }
    \label{fig:phase_resolved}
\end{figure*}
To investigate the underlying cause of PA variation over time, we conducted a pulse phase-resolved analysis. To ensure sufficient statistics for phase-resolved analysis, we regrouped the data into a single time interval for Obs. 1 and three intervals for Obs. 2. The pulse was divided into 16 bins, with the results presented in Fig.~\ref{fig:phase_resolved}. The two observations exhibit markedly different pulse profiles, while the shapes of the pulse profiles in the three intervals of Obs. 2 remain consistent. In the first interval (Obs. 1), the PD for all phase bins is below 18\%. In contrast, for the three intervals in Obs. 2, the PD reaches up to 25\%.

Although the PD exhibits a relatively complex variation as a function of pulse phase, the evolution of the PA follows a simpler pattern. To model the phase-dependent behavior of the PA, we utilized the RVM  \citep{Radhakrishnan_RVM, Poutanen_RVM}. Under the assumption that the radiation primarily escapes via the ordinary mode (O-mode), the PA is described by the following equation:

\begin{equation}
\label{equ:RVM}
\tan(\text{PA} - \chi_{\mathrm{p}}) = \frac{-\sin{\theta}\sin(\phi - \phi_{0})}{\sin{i_{\mathrm{p}}}\cos{\theta} - \cos{i_{\mathrm{p}}}\sin{\theta}\cos(\phi - \phi_{0})},
\end{equation}

where \(i_{\rm p} \in (0\degr, 180\degr)\) is the inclination angle between the pulsar's spin axis and the observer's line of sight, \(\theta_{\rm p} \in (0\degr, 90\degr)\) represents the magnetic obliquity (i.e., the angle between the magnetic dipole and the spin axis), \(\chi_{\mathrm{p}} \in (-90\degr, 90\degr)\) is the position angle of the pulsar's spin axis, \(\phi\) is the pulse phase, and \(\phi_{\mathrm 0} \in (0,1)\) denotes the phase at which the magnetic pole is closest to the observer.

The RVM has been applied to model the phase-dependent PA variations in several X-ray accreting pulsars to determine their geometric parameters \citep{Doroshenko_etal_2022_herx-1, Tsygankov_etal_2022_cenx-3, Mushtukov_xpeisei, Tsygankov_groj1008, Marshall_4U1627, Malacaria_exo2030, Long_0535, Victor_etal_2023, Zhao_herx-1, Heyl_herX-1, SMCX-1_Forsblom, SwiftJ0243_Poutanen}. To fit the RVM to the observed PA, we follow the method outlined by \citet{SwiftJ0243_Poutanen, SMCX-1_Forsblom}, as the PA is not normally distributed. We use the probability density function of the PA, \(\psi\), derived by \citet{Naghizadeh1993}:

\begin{equation} 
\label{eq:PA_G}
G(\psi) = \frac{1}{\sqrt{\pi}} 
\left\{  \frac{1}{\sqrt{\pi}}  + 
\eta \, {\rm e}^{\eta^2} 
\left[ 1 + {\rm erf}(\eta) \right]
\right\} {\rm e}^{-p_0^2/2}.
\end{equation}

Here, \(p_0 = \sqrt{q^2 + u^2}/\sigma_{\rm p}\) represents the measured PD in units of its uncertainty, \(\eta = p_0 \cos[2(\chi - \chi_0)]/\sqrt{2}\), where \(\chi_0 = \frac{1}{2} \arctan(u/q)\) is the central PA derived from the Stokes parameters, \(\chi\) is the PA predicted by the RVM model (given by Eq.~\ref{equ:RVM}), and \(\mathrm{erf}\) denotes the error function.

The fitting procedure was conducted using the \texttt{EMCEE} package, an affine-invariant Markov Chain Monte Carlo (MCMC) ensemble sampler implemented in \texttt{Python} \citep{Foreman-Mackey_emcee}. The best-fitting parameters are summarized in Table~\ref{tab:RVM fitting}, and the posterior distributions are shown in Fig.~\ref{fig:corner_plot_rvm}. Interestingly, the RVM parameters for Obs. 1 differ significantly from those of Obs. 2, as previously reported by \citet{Victor_etal_2023}. Given the relatively short time interval ($\sim$ 20 days) between the two observations, such a substantial change in geometry is unexpected. More intriguingly, we discovered a continuous evolution of the RVM parameters during Obs. 2, which may indicate a complex, ongoing evolution of the different polarized components contributing to the total emission from the source.
We employ a two-polarized component model as described in \citet{Victor_etal_2023}, assuming the presence of an additional polarized component that remains constant with respect to pulse phase, alongside the RVM-predicted polarized component. We applied a two-polarized components model to the four intervals. In this model, the observed Stokes parameters can be expressed as the sum of the absolute Stokes parameters for the variable and constant components:

\begin{eqnarray}  
I(\phi) &=& I_{\mathrm{c}} + I_{\mathrm{p}}(\phi) , \nonumber \\
Q(\phi) &=& Q_{\mathrm{c}} + Q_{\mathrm{p}}(\phi) , \\
U(\phi) &=& U_{\mathrm{c}} + U_{\mathrm{p}}(\phi) .  \nonumber
\end{eqnarray} 

The subscripts \textit{c} and \textit{p} represent the constant and variable components, respectively. In this analysis, the Stokes parameters are normalized to the average flux, and the Stokes parameters of the constant component are assumed to remain invariant with pulse phase. The Stokes parameters \(Q\) and \(U\) are related to the polarized flux and PA as follows:

\begin{eqnarray}  
Q_{\mathrm{p}}(\phi) &=& PD_{\mathrm{p}}(\phi) \, I_{\mathrm{p}}(\phi) \cos[2\chi(\phi)] = PF_{\mathrm{p}}(\phi) \cos[2\chi(\phi)], \nonumber \\
U_{\mathrm{p}}(\phi) &=& PD_{\mathrm{p}}(\phi) \, I_{\mathrm{p}}(\phi) \sin[2\chi(\phi)] = PF_{\mathrm{p}}(\phi) \sin[2\chi(\phi)],  \\
Q_{\mathrm{c}} &=& PD_{\mathrm{c}} \, I_{\mathrm{c}} \cos[2\chi_{\mathrm{c}}] = PF_{\mathrm{c}} \cos[2\chi_{\mathrm{c}}] , \nonumber \\
U_{\mathrm{c}} &=& PD_{\mathrm{c}} \, I_{\mathrm{c}} \sin[2\chi_{\mathrm{c}}] = PF_{\mathrm{c}} \sin[2\chi_{\mathrm{c}}] . \nonumber
\end{eqnarray} 

Here, \(PD_{\mathrm{p}}(\phi)\) and \(PD_{\mathrm{c}}\) represent the PD of the variable and constant components, respectively. \(\chi_{\mathrm{c}}\) denotes the PA of the constant component, while \(\chi(\phi)\) corresponds to the PA of variable component. \(PF_{\mathrm{p}}(\phi)\) and \(PF_{\mathrm{c}}\) represent the polarized flux in units of the average flux. We fitted the observed Stokes parameters \textit{I, Q, U}, assuming uniform priors for all parameters except for \(i_{\rm p}\) and \(\theta_{\rm p}\), for which flat priors were used for the cosine of the angles. The likelihood was calculated using \(\chi^2\) statistics for the Stokes parameters \(Q\) and \(U\) \citep{Victor_etal_2023}. In this fitting, the prior ranges were set as follows: \(i_{\rm p} \in (0\degr, 180\degr)\), \(\theta_{\rm p} \in (0\degr, 90\degr)\), \(\chi_{\mathrm{p}} \in (-90\degr, 90\degr)\), \(\phi_{\mathrm{p}} \in (0, 1)\), \(PD_{\mathrm{c}} \in (0, 1)\), \(PD_{\mathrm{p}} \in (0, 1)\), and \(I_{\mathrm{c}} \leq I(\phi)_{\min}\).

\begin{figure}[tb]
    \centering
    \includegraphics[width=0.45\textwidth] {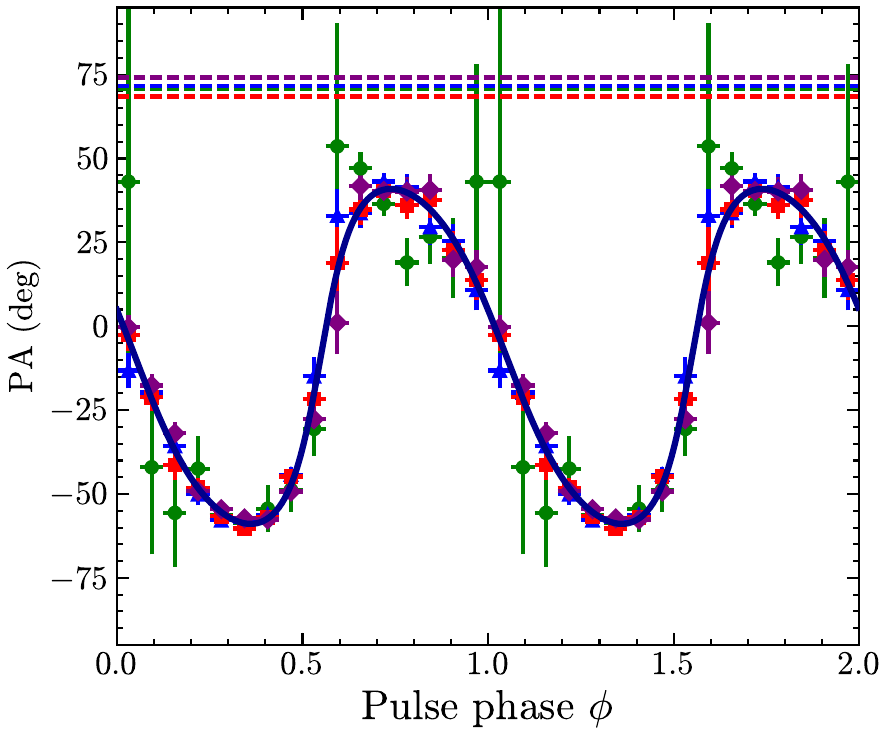}    
    \caption{The PA for the variable components is indicated by colored points, while the constant components are represented by dashed lines. The dark blue solid line represents the RVM curve after subtracting the constant components. The color coding is the same as in Fig.~\ref{fig:phase_resolved}. }
    \label{fig:phase_resolved_tc}
\end{figure}

\begin{figure*}[tb]
    \centering
    \includegraphics[width=1.0\textwidth] {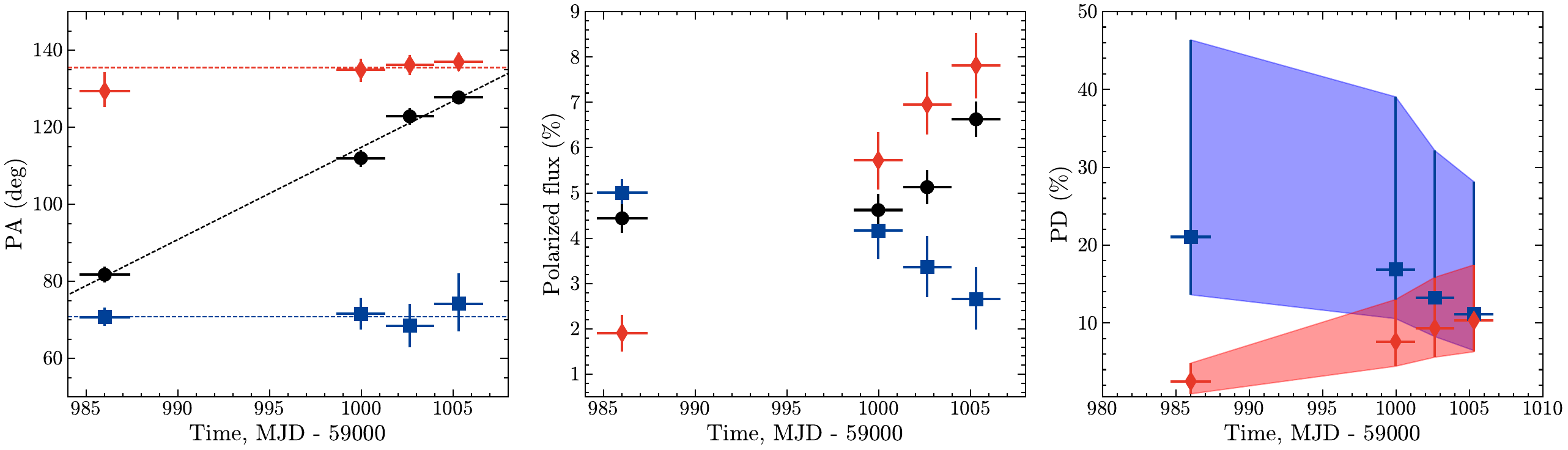}

    \caption{Left panel: the pulse phase-averaged PA variations over time, where black dots represent the total PA, red diamonds denote the PA of the variable component, and blue squares indicate the PA of the constant component. Middle and right panels: polarized flux and PD variations over time, using the same markers as in the left panel, respectively.    }
    \label{fig:time_resolved_tc}
\end{figure*}

Through this fitting process, we are able to constrain the four RVM parameters and the Stokes parameters \(Q_{\mathrm{c}}\) and \(U_{\mathrm{c}}\) for each interval as illustrated in Fig.~\ref{fig:corner_plot_IQU}. From these, we calculate the polarized flux \(PF_{\mathrm{c}}\) and the PA \(\chi_{\mathrm{c}}\) distributions from the MCMC chains for each interval with the formula of Eq.~\ref{eq:PD_c},
\begin{align}  
\label{eq:PD_c}
PF_{\mathrm{c}}  &= \sqrt{Q_{\mathrm{c}}^2 + U_{\mathrm{c}}^2},  \nonumber  \\
\chi_{\mathrm{c}}  &= \frac{1}{2} \arctan{\left(\frac{U_{\mathrm{c}}}{Q_{\mathrm{c}}}\right)}, 
\end{align}
The \(PF_{\mathrm{c},i}\) values for \(i = 1, 2, 3,\) and 4 are \(5.0 \pm 0.3\%\), \(4.2 \pm 0.6\%\), \(3.4 \pm 0.7\%\), and \(2.7 \pm 0.7\%\), as shown in the Fig.~\ref{fig:corner_plot_constant}, respectively. Notably, \(\chi_{\mathrm{c}}\) remains approximately 70\degr across all intervals, as shown in Figs.~\ref{fig:phase_resolved_tc} and \ref{fig:time_resolved_tc}. By subtracting the Stokes parameters \(Q_{\mathrm{c}}\) and \(U_{\mathrm{c}}\) from the total Stokes values in each phase bin, we compute the PA of the variable component using the following expression: 
\begin{align}  
\label{eq:PA_p}
\chi(\phi)  &= \frac{1}{2} \arctan{\left(\frac{{U_{\mathrm{p}}(\phi)}}{{Q_{\mathrm{p}}}(\phi)}\right)}.
\end{align}
As illustrated in Fig.~\ref{fig:phase_resolved_tc}, the PA variations with pulse phase for the variable component across all four intervals can be modeled using a single set of RVM parameters.

\begin{table*}[tb]
    \centering
    \resizebox{0.8\textwidth}{!}{%
    \begin{tabular}{ccccccc} \hline
        Observation  & Interval & MJD & $i_{\rm p}$ ($^{\circ}$) & $\theta$ ($^{\circ}$) & $\chi_{\rm p}$ ($^{\circ}$) & $\phi$/2$\pi$ \\ \hline
        02250401  & Interval 1 & 59984.647-59987.396 & $54^{+10}_{-12}$ & $32^{+4}_{-6}$ & $83.4\pm0.9$ & $0.565^{+0.009}_{-0.008}$\\ 
 
     02250501  & Interval 2 & 59998.657-60001.311 & $101\pm3$ & $56\pm1$ & $-5.9\pm0.8$ & $0.039\pm0.005$\\
         & Interval 3 & 60001.331-60003.966 & $100^{+3}_{-2}$ & $53.6\pm1$ & $-7.2^{+0.8}_{-0.7}$ & $0.052\pm0.004$\\
         & Interval 4 & 60003.973-60006.661 & $106^{+3}_{-2}$ & $51.5^{+1.1}_{-1.2}$ & $-6.1\pm0.8$ & $0.060\pm0.004$\\
         \hline
    \end{tabular}%
    }
   \caption{RVM fitting parameters.
   }
   \label{tab:RVM fitting}
\end{table*}

Since the motivation of this analysis is to investigate the underlying cause of the PA rotation observed in panel (c) of Fig.~\ref{fig:time_resolved}, we also calculated the phase-averaged PA ($\chi$) and polarized flux ($PF_{\rm p}$) of variable component for each interval using the formula of Eq.~\ref{eq:PFp}. 
\begin{align}  
\label{eq:PFp}
PF_{\mathrm{p}}  &= \frac{\sqrt{(\sum{{Q_{\mathrm{p}}(\phi)}})^2 + (\sum{{U_{\mathrm{p}}(\phi)})^2}}}{{N_{\mathrm{bin}}}},  \nonumber  \\
\chi  &= \frac{1}{2} \arctan{\left(\frac{\sum{U_{\mathrm{p}}(\phi)}}{\sum{Q_{\mathrm{p}}}(\phi)}\right)}.
\end{align}
Where $N_{\mathrm{bin}}$ represents the number of phase bins.

As shown in the left panel of Fig.~\ref{fig:time_resolved_tc} and Fig.~\ref{fig:corner_plot_variable}, the phase-averaged PAs of variable components also roughly remain unchanged over time. However, the polarized flux of both components changes over time. During the first interval, the polarized flux of the constant component is significantly higher than that of the variable component but decreases with time/luminosity. In contrast, the polarized flux of the variable component increases with time/luminosity, ranging from 2\% to 8\%. Hence, the observed PA variation between intervals is driven by the competing contributions of the polarized flux from the constant and variable components. 

Furthermore, it is important to discuss the constraints on \(I_{\mathrm{c}}\) and \(PD_{\mathrm{c}}\), as they offer valuable insights into the potential origin of the constant component. The limits on \(I_{\mathrm{c}}\) are derived from the conditions \(PD_{\mathrm{c}} \in (0, 1)\), \(PD_{\mathrm{p}} \in (0, 1)\), and \(I_{\mathrm{c}} \leq I(\phi)_{\min}\). Based on this, the limits on \(I_{\mathrm{c},i}\) are approximately $[0.11, 0.37]$, $[0.11, 0.39]$, $[0.11, 0.40]$, and $[0.10, 0.38]$ for \(i = 1, 2, 3,\) and 4, respectively, as shown in the Fig.~\ref{fig:corner_plot_IQU}. \
Since \(I_{\mathrm{c}}\) is inversely proportional to \(PD_{\mathrm{c}}\), as shown in Eq.~\ref{eq:PD}, the corresponding ranges for \(PD_{\mathrm{c},i}\) can be calculated from the formula:
\begin{align}  
\label{eq:PD}
PD_{\mathrm{c}}  &= \frac{\sqrt{Q_{\mathrm{c}}^2 + U_{\mathrm{c}}^2}}{I_{\mathrm{c}}}.
\end{align}

The PD of constant component are $[14, 46]$, $[11, 39]$, $[8, 32]$, and $[6, 28]$ percent. Using the median values (24\%-25\%), \(PD_{\mathrm{c},i}\) is approximately $21\%$, $17\%$, $13\%$, and $11\%$ as shown in the Fig.~\ref{fig:corner_plot_constant}. Assuming \(PD_{\mathrm{c},i} \leq 25\%\), the minimum of \(I_{\mathrm{c},i}\) lies within the range of 11\% to 20\%, depending on the specific interval.

With these constraints on \(I_{\mathrm{c}}\), we can similarly limit \(I_{\mathrm{p}}\). By combining the values of \(PF_{\mathrm{p}}\) and \(I_{\mathrm{p}}\), we can calculate \(PD_{\mathrm{p}}\) with following expression:
\begin{align}  
\label{eq:PDp}
PD_{\mathrm{p}}  &= \frac{\sqrt{\left(\sum{Q_{\mathrm{p}}(\phi)}\right)^2 + \left(\sum{U_{\mathrm{p}}(\phi)}\right)^2}}{\sum{I_{\mathrm{p}}(\phi)}}.
\end{align}

As shown in the right panel of Fig.~\ref{fig:time_resolved_tc} and Fig.~\ref{fig:corner_plot_variable}, \(PD_{\mathrm{p}}\) lies between 2\% and 13\%, which is generally lower than \(PD_{\mathrm{c}}\). The PD of the constant component remains relatively high, but with significant uncertainties due to the broad range of \(I_{\mathrm{c}}\). In contrast, the variable component has a narrower PD range. This is because PD is inversely related to \(I\). For \(I_{\mathrm{c}}\), the range is roughly [0.10, 0.40], while for \(I_{\mathrm{p}}\), it is [0.40, 0.90]. As a result, the range for \(PD_{\mathrm{p}}\) is approximately [1.1$PF_{\mathrm{min}}$, 2.5$PF_{\mathrm{max}}$], whereas for \(PD_{\mathrm{c}}\), it extends from [2.5$PF_{\mathrm{min}}$, 10$PF_{\mathrm{max}}$], making it more sensitive to changes in \(I_{\mathrm{c}}\).

\section{Discussion and summary}
\label{sec:sec4}

In this paper, we performed a polarimetric analysis of the transient XRP RX~J0440 using two IXPE observations taken during its 2022 outburst.
\citealt{Victor_etal_2023} previously reported that as the luminosity of the XRP increases, an additional component emerges in the polarized radiation, exhibiting a constant polarization throughout the pulsation period.
However, analysis performed by \citealt{Victor_etal_2023} was limited to studying two luminosity states of the XRP. 
By reanalyzing the same data, we were able to investigate the evolution of the polarization properties in RX~J0440 with higher time resolution 
compared to previous analysis \citep{Victor_etal_2023}.
This allowed us to trace how the component with constant polarization varies as a function of time and X-ray luminosity with greater details.

We have discovered that the phase-averaged PA in RX~J0440 undergoes a gradual rotation, shifting from  $\sim 80\,{\rm deg}$ at the highest observed luminosity to $\sim 130\,{\rm deg}$ at the lowest luminosity state captured by IXPE observations (see Fig.\,\ref{fig:time_resolved}c).
The rotation of the PA and the decrease in luminosity are accompanied by an increase in the PD (see Fig.\,\ref{fig:time_resolved}b) and a reduction in the RMS pulsed fraction (see Fig.\,\ref{fig:time_resolved}a).

Decomposing the total X-ray polarization signal into pulsed and constant components reveals that the observed changes in polarization properties with luminosity can be attributed to the evolving contribution of these two components (see middle panel in Fig.\,\ref{fig:time_resolved_tc}).
According to our analysis, the relative contribution of the constant polarization component increases with the luminosity of RX~J0440 (blue squared in Fig.\,\ref{fig:time_resolved_tc}), while the contribution of the pulsed component decreases (red diamonds in Fig.\,\ref{fig:time_resolved_tc}). 
Notably, the phase-averaged PA of both components remains largely unchanged in time (and at different mass accretion rates). 
Based on the available data and the method used for processing, we can obtain a reliable estimate of the polarized flux of both components. 
While the PD estimate is somewhat less precise, we observe the following trends (see right panel in Fig.\,\ref{fig:time_resolved_tc}): 
(i) the PD of the pulsating component increases with decreasing luminosity, rising from $\sim$ $3\%$ in the bright state to approximately $8\%$ in the low-luminosity state, 
(ii) the PD of the constant component is significantly higher than that of the pulsating component, with an estimate exceeding $15\%$ in the bright state, and 
(iii) it is possible that the PD of the constant component decreases with luminosity, although this requires further confirmation.

The pulsating component is likely associated with the radiation emitted from the surface of the NS, as discussed by \citealt{Victor_etal_2023}. 
However, the origin of the constant polarization component remains unclear. 
Notably, the phase-averaged PA of the pulsating and constant components differs by 60--70 deg (see left panel in Fig.\,\ref{fig:time_resolved_tc}). 

This suggests that the pulsating and constant components have different origins. 
We assume that the phase-resolved PA of the pulsating component follows the RVM, which indicates that this component forms within the adiabatic radius of the magnetized NS \citep[in XRPs, this radius is expected to be $\lesssim 10R_{\rm NS}$, see, e.g.,][]{2002PhRvD..66b3002H, 2015MNRAS.454.3254T}. 
In this region, the polarization plane rotates
following local direction of magnetic field 
as the radiation travels through a NS magnetosphere, reaching up to the adiabatic radius. 
Our assumption that the dipole magnetic field component dominates allows the application of the RVM: the linearly polarized component of the X-ray flux emerging from the adiabatic radius aligns either parallel or perpendicular to the direction of the magnetic field.
In contrast, the phase-averaged PA of the constant component differs from that of the pulsating component. 
In our analysis, we assumed that the additional polarization component does not vary in PA throughout the pulsation period. 
Under this assumption, the component likely originates from a region outside the adiabatic radius.

While we cannot entirely rule out the possibility of pulsations in this component, if its PA remains unchanged during the NS spin period, this could impose constraints on the potential interpretations of its origin.
Among the potential candidates for explaining the additional polarization component, we highlight two main mechanisms: 
(1) scattering of X-rays by outflows launched from the inner regions of the accretion disk (see discussion and references in Sect. 4.2 in \citealt{Victor_etal_2023}), and 
(2) X-ray scattering by the magnetospheric accretion flow \citep{2017MNRAS.467.1202M,2024MNRAS.529.1571F}. 
In both scenarios, the contribution of scattered radiation is expected to increase with luminosity, as both the mass outflow rate from the disk and the optical thickness of the magnetospheric accretion flow rise with luminosity.
However, it is important to consider that if the scattering is attributed to outflows from the disk, we must address why the intensity of these outflows varies significantly with relatively small changes in luminosity (notably, the luminosity in the IXPE data changes within a factor of three only). 
On the other hand, while the hypothesis of scattering by the magnetospheric outflow can account for the rapid changes in the contribution of the additional component with luminosity (at a luminosity of $\sim 10^{36}\,\ergs$, approximately $1-2\%$ of photons are scattered by the flow, while at $\sim 10^{37}\,\ergs$, the fraction of scattered photons can be $\sim 10\%$ depending on X-ray beam pattern, Mushtukov et al., in prep.), it still requires an explanation for why this component does not exhibit pulsations in its PA.

\begin{acknowledgements}
We acknowledge funding support from the National Natural Science Foundation of China under grants Nos.\ 12122306, 12025301, \& 12103027, and the Strategic Priority Research Program of the Chinese Academy of Sciences.
AAM thanks UKRI Stephen Hawking fellowship.

\end{acknowledgements}
\bibliographystyle{yahapj}

\bibliography{ref}

\begin{appendix}
\section{Posterior distributions of single-component RVM parameters.}
\begin{figure}[h!]
    \centering
    \includegraphics[width=0.45\textwidth] {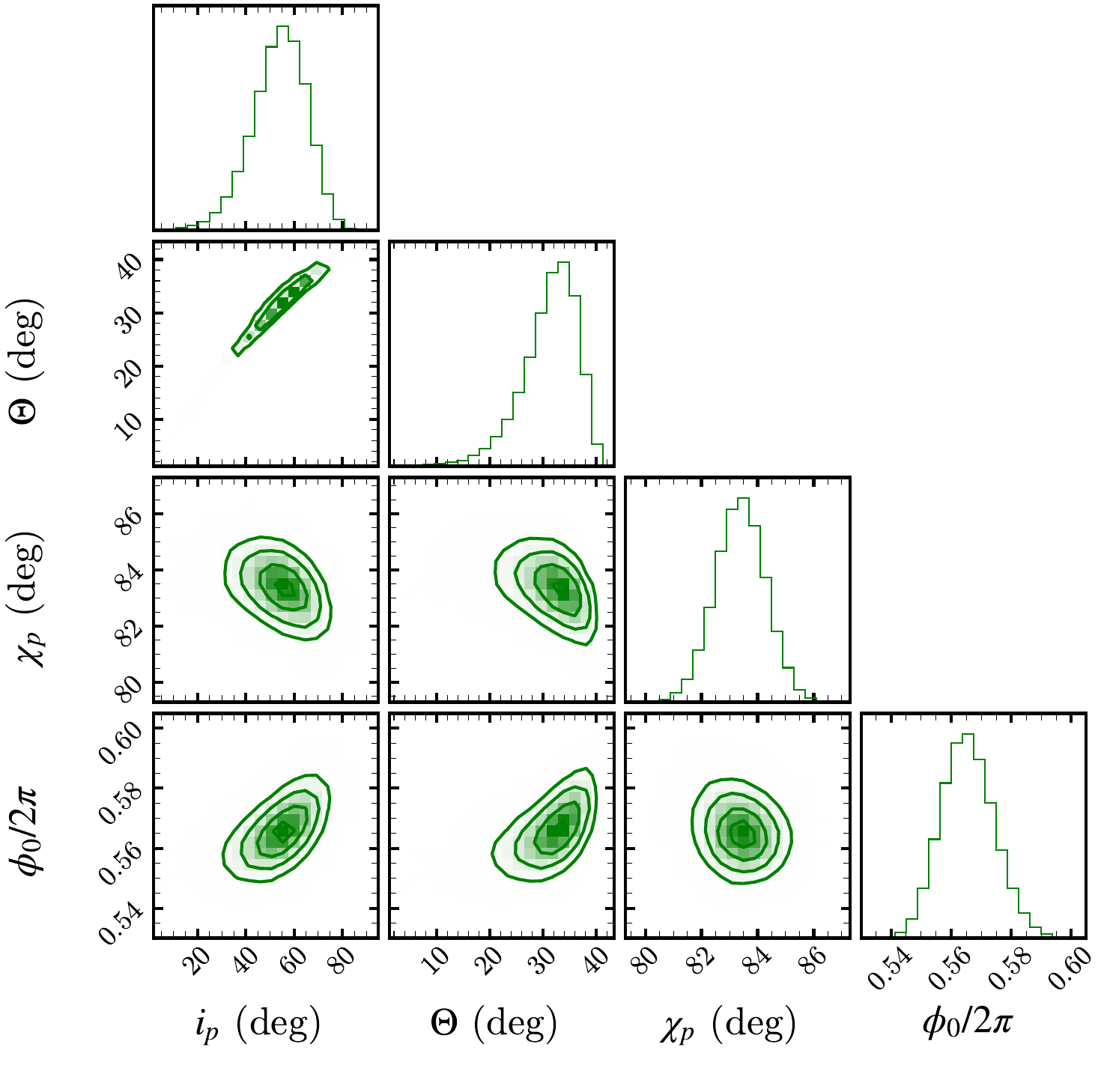}
    \includegraphics[width=0.45\textwidth] {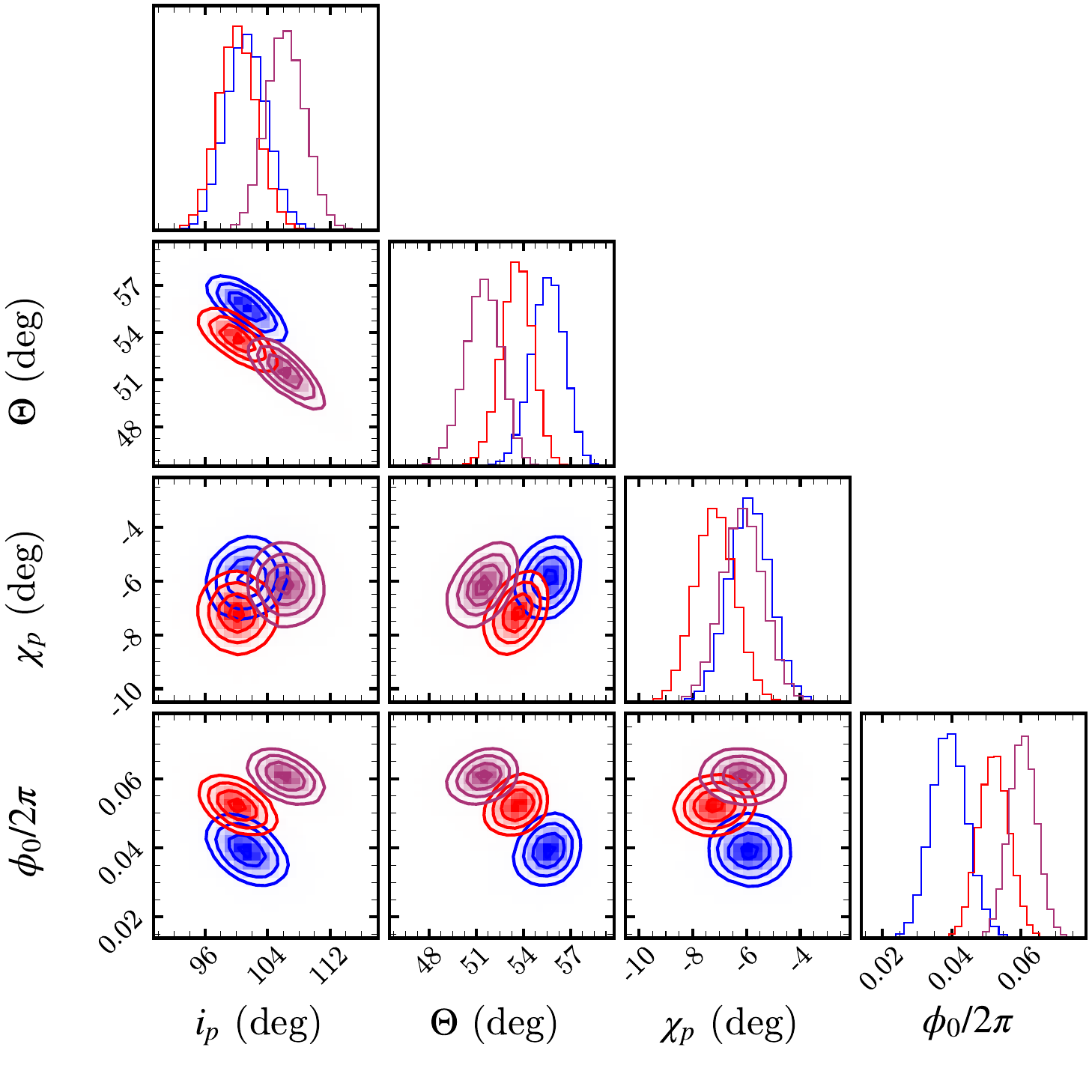}
    \caption{Corner plots of the posterior distributions for the single-component RVM parameters. The distributions for intervals 1, 2, 3, and 4 are color-coded in green, blue, coral red, and purple, respectively.
}
    \label{fig:corner_plot_rvm}
\end{figure}
\section{Posterior distributions of two-component RVM parameters.}
\begin{figure*}[h!]
    \centering
    \includegraphics[width=1.0\textwidth] {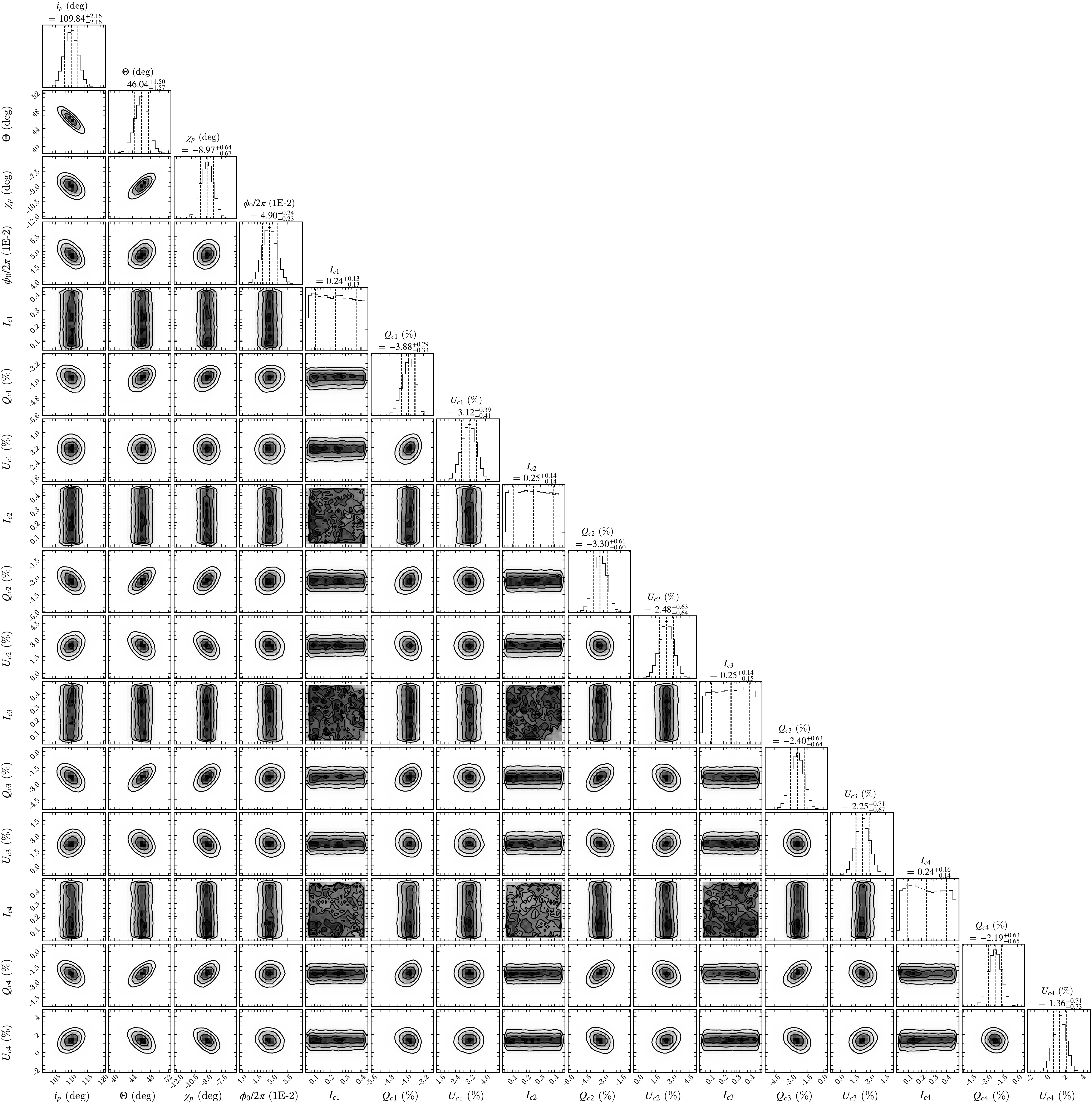}
    \caption{Corner plot of the posterior distribution for parameters of the two-component RVM model. The Stokes parameters $Q_{\mathrm c \ i}$ and $U_{\mathrm c \ i}$, i = 1, 2, 3 and 4 are expressed as percentages of the averaged flux,  $I_{\mathrm c \ i}$  given as a fraction of the average flux.}
    \label{fig:corner_plot_IQU}
\end{figure*}
\begin{figure*}[h!]
    \centering
    \includegraphics[width=1.0\textwidth] {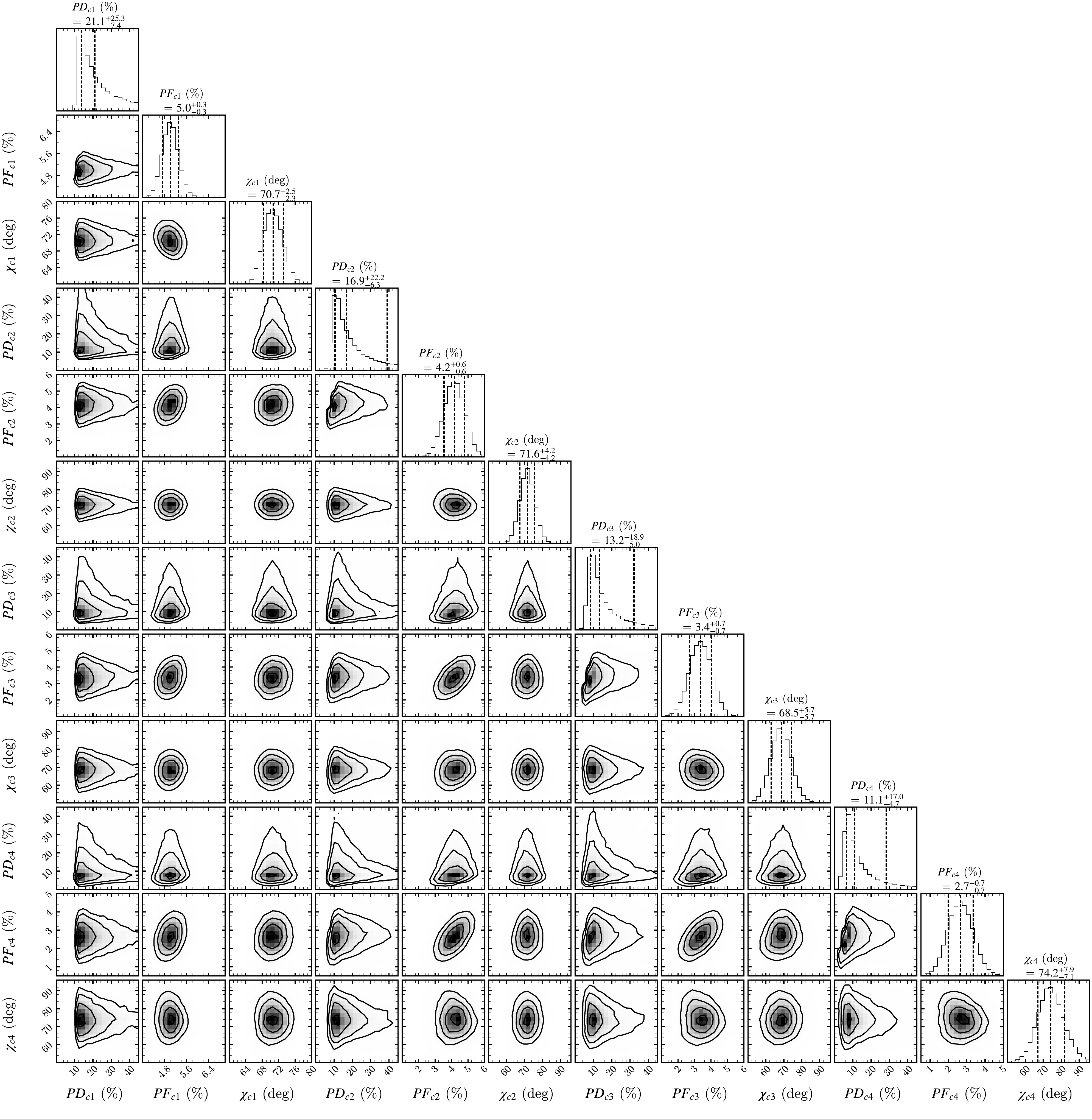}
    \caption{Corner plot of the posterior distribution for parameters of the two-polarization component model. $PF_{\mathrm c \ i}$, i = 1, 2, 3 and 4 are the polarized flux of the constant component in unit of averaged flux for four individual time intervals. $\chi_{\mathrm c \ i}$ are its polarization angle.}
    \label{fig:corner_plot_constant}
\end{figure*}
\begin{figure*}[h!]
    \centering
    \includegraphics[width=1.0\textwidth] {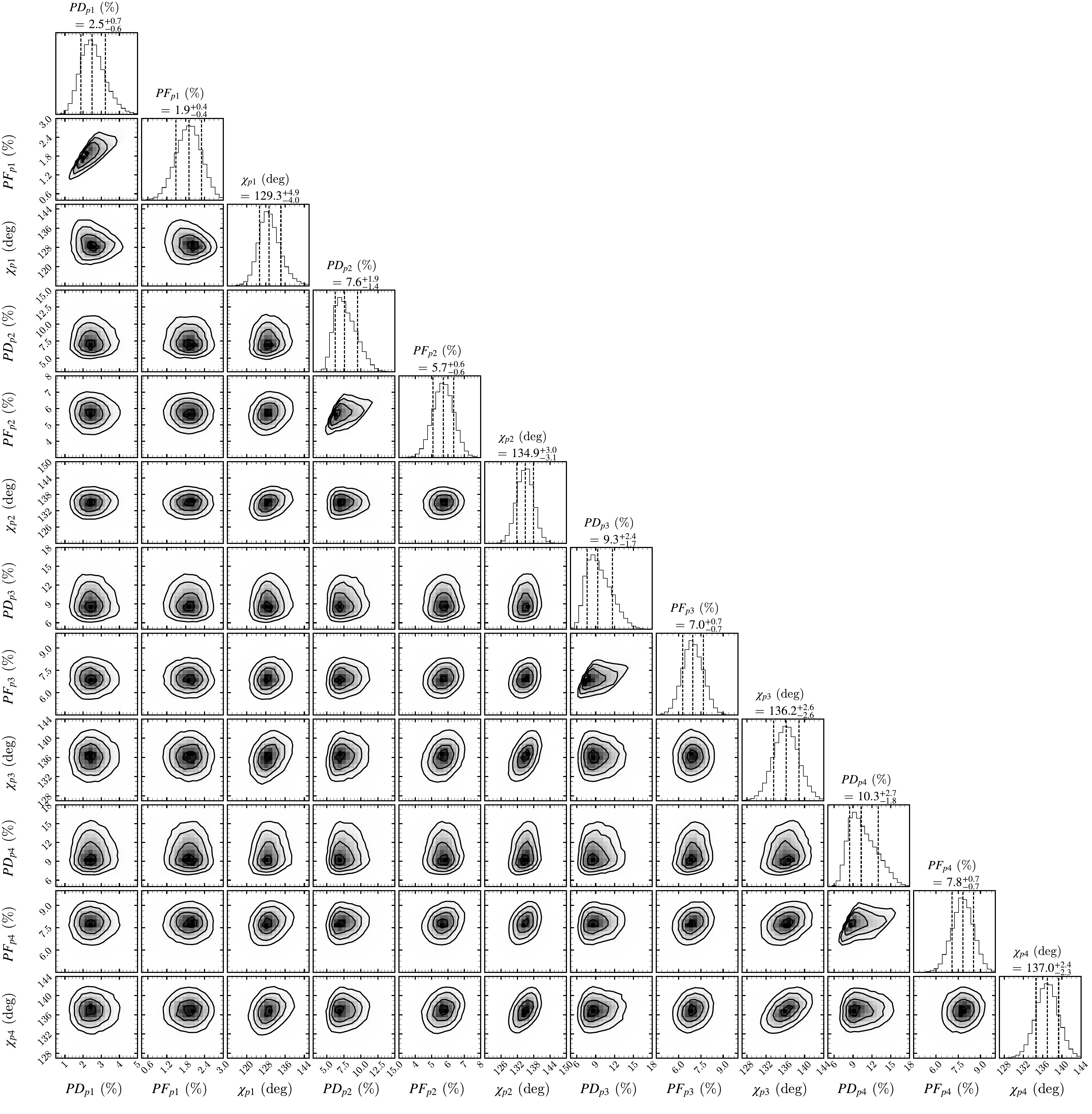}
    \caption{Corner plot of the posterior distribution for parameters of the two-polarization component model. $PF_{\mathrm p, \ i}$, i = 1, 2, 3 and 4 are the polarized flux of the pulsed component in unit of averaged flux for four individual time intervals. $\chi_{\mathrm p, \ i}$ are its polarization angle.}
    \label{fig:corner_plot_variable}
\end{figure*}
\end{appendix}
%
%

\end{document}